\documentstyle[11pt]{article}

\title{The method of vacuum vectors in the theory of \goodbreak
Yang -- Baxter equation}
\author{Igor G. Korepanov\thanks{South Ural State University,
76 Lenin ave., Chelyabinsk 454080, Russia.
E-mail address: kig@susu.ac.ru}}
\date{Translated from the book: Applied Problems in Calculus, Publishing
House of Chelyabinsk Polytechnical Institute, Chelyabinsk, Russia, 1986,
pages 39--48}

\long\def\oip#1{\par\medskip\par{\bf#1}\vadjust{\nobreak}\ }

\def\la{\lambda}
\def\bz{\hbox{\Huge 0}}
\def\hm{]}

\begin{document}
\maketitle

\begin{abstract}
In modern terminology, this is the first published paper where the
solutions of Yang--Baxter equation ``at roots of unity'' were analyzed
and shown to be related to algebraic curves of genus $>1$. They are
also known now to be connected with the ``chiral Potts model''.
The paper's abstract as written in 1986 reads: ``Vacuum vectors
of an $L$-operator form a holomorphic bundle over the vacuum curve
of that operator. These notions, as well as the theory of commutation
relations of the 6-vertex model, are used in this work for constructing
solutions of the Yang--Baxter equation that do not possess a spectral
parameter of traditional type''.
\end{abstract}

In this paper the properties of solutions of Yang--Baxter equation
are studied connected with the existence of the so-called vacuum vectors
of those solutions. In \S1 general definitions and theorems are stated
concerning vacuum vectors and their relation to Yang--Baxter equation.
In \S2 and~3 the problem of existence of some new solutions to
Yang--Baxter equation in solved on the base of those definitions and
theorems. These solutions are intimately connected with the well-known
``6-vertex model'' and at the same time possess remarkable new properties
considered in~\S4.

\section*{\S1. Vacuum vectors and vacuum curves}

Let $S^Q$ and $S^A$ be two finite-dimensional complex linear spaces ($S^Q$
is called ``quantum space'', and $S^A$ is ``auxiliary'', see~[1]); and
let $L$ be a linear operator acting in $S^Q\otimes S^A$. We will denote
vectors from $S^Q$ by letters $U,V,\ldots$, and vectors from $S^A$ by
letters $X,Y,\ldots$. Let the relation
$$
L(U\otimes X)=V\otimes Y
\eqno(1)
$$
hold.

\oip{Definition 1.} The {\em vacuum variety\/} $\Gamma_L$ of an operator $L$
is the set of pairs $(U,V)$, $U\ne 0$, $V\ne 0$, for which the relation (1)
is satisfied with some $X$ and $Y$,\quad $X\ne 0$, taken to within the
equivalence $(U,V) \sim (t_1U,t_2V)$, where $t_1$ and $t_2$ are some
nonzero numbers. Any (including zero) vector~$X$ corresponding according
to~(1) to a given point $z\in \Gamma_L$ is called {\em vacuum vector},
while any vector~$Y$ --- {\em co-vacuum vector}. The vacuum variety is
an algebraic variety, and a point~$z$ must be assumed to enter it with
the multiplicity equal to the dimension of the space of vacuum vectors in it.

\oip{Definition 2.} An operator $L$ acting in $S^Q\otimes S^A$ is called
{\em equivalent\/} to an operator~$M$ acting in $S^Q\otimes S^B$ if
an isomorphism $R\colon\; S^A\to S^B$ exists such that $RL=MR$.
Similarly, two families of operators $L(\la)$ and $M(\la)$
($\la$ is any parameter) are called equivalent if an isomorphism $R$
exists such that $RL(\la)=M(\la)R$ for every~$\la$.
In these cases we will say that $R$ {\em performs\/} such equivalence.

\oip{Remark.} Here we identify the operator $R$ with the operator
${\bf 1}\otimes R$ acting from $S^Q\otimes S^A$ to $S^Q\otimes S^B$.
We also treat similarly other operators when needed.

\medskip

The following lemma, although being evident, plays the fundamental role
in the sequel.

\oip{Lemma 1.} The vacuum varieties of equivalent operators coincide.

\medskip

In the rest of this paper we consider only the case $\dim S^Q=2$.
We can then put in formula~(1)\quad
$U=\pmatrix{u\cr 1}$, $V=\pmatrix{v\cr 1}$,
where $u$ and $v$ are complex numbers or~$\infty$. Denote $\dim S^A$
as~$N$.

\oip{Theorem 1}[2]. The vacuum curve $\Gamma_L$ of a generic operator~$L$
is a smooth irreducible algebraic curve given by equation of the kind
$$
P_L(u,v)=\sum_{i,j=1}^N a_{ij}u^i v^j =0.
\eqno(2)
$$
The vacuum (co-vacuum) vectors form a one-dimensional holomorphic bundle
of degree $N^2-N$ over~$\Gamma_L$.

\medskip

Degenerate cases are also important, for example the case when
$P_L(u,v)=[p(u,v)]^l$. In such case the vacuum (co-vacuum) vectors form an
$l$-dimensional holomorphic bundle over the curve given by the polynomial
$p(u,v)$~[2].

Introduce now two auxiliary spaces $S_1^A$ and $S_2^A$, and let
$S^A=S_1^A\otimes S_2^A$. Consider operators $L_1$ and $L_2$ acting
respectively in $S^Q\otimes S_1^A$ and $S^Q\otimes S_2^A$, and the
operator $L_1L_2$ acting in $S^Q\otimes S^A$.

\oip{Definition 3.} {\em Composition\/} of vacuum curves $\Gamma_1$ and
$\Gamma_2$ given by equations $P_1(u,v)=0$ and $P_2(u,v)=0$ of the form~(2)
is the curve~$\Gamma$ given by equation $P(u,v)=0$ where $P(u,v)$ is the
resultant~[3] of $P_1(w,v)$ and $P_2(u,w)$ considered as polynomials in~$w$.

\medskip

Thus, $\Gamma$ is the set of points $u,v$, taken with proper multiplicities,
for which a $w$ exists such that $(u,w)\in\Gamma_2$ and $(w,v)\in\Gamma_1$.

\oip{Theorem 2}[2]. The vacuum curve $\Gamma$ of the operator $L_1L_2$ is
the composition of vacuum curves $\Gamma_1$ and $\Gamma_2$ of operators
$L_1$ and~$L_2$.

\medskip

We are interested in the equivalence of operators of the form $L_1L_2$ and
$L_2L_1$, that is in solutions of Yang--Baxter equation $RL_1L_2=L_2L_1R$.
This motivates, in light of Lemma~1 and Theorem~2, the following

\oip{Definition 4.} Vacuum curves $\Gamma_1$ and $\Gamma_2$ are called
{\em commuting\/} if the composition of $\Gamma_1$ and $\Gamma_2$ coincides
with the composition of $\Gamma_2$ and $\Gamma_1$.

\section*{\S2. Formulation of the main theorem}

Recall that $\dim S^Q=2$, and let $\dim S_1^Q=\dim S_2^Q=3$. Considers two
one-parametric families of operators:
$$
L_i(\la)=\pmatrix{A_i(\la) & B_i(\la) \cr
                      C_i(\la) & D_i(\la)}
\eqno(3)
$$
($i=1,2$), where $A_i(\la),\ldots,D_i(\la)$ act in spaces $S_i^A$,
realize two representations of the commutation relations of 6-vertex model
[1,4--7] with parameter $\eta=\pi/3$, and are expressed, in the propers
bases, by formulas
$$
A_i(\la)=a_i\pmatrix{\sin(\la+\rho_i) && \cr
& \sin(\la+\rho_i-{2\pi\over 3}) & \cr & & \sin(\la+\rho_i-
{4\pi\over 3})},
\eqno(4)
$$
$$
D_i(\la)=d_i\pmatrix{\sin(\la+\sigma_i-{4\pi\over 3}) && \cr
& \sin(\la+\sigma_i-{2\pi\over 3}) & \cr & & \sin(\la+\sigma_i)},
\eqno(5),
$$
$$
B_i(\la)=C_i^{\rm T}(\la)=
\pmatrix{0 && b_{13}^{(i)} \cr b_{21}^{(i)} & 0 & \cr & b_{32}^{(i)} & 0},
\eqno(6)
$$
$$
(b_{k+1,k}^{(i)})^2-(b_{k,k-1}^{(i)})^2=a_i d_i \sin{2\pi\over 3}
\sin(\rho_i-\sigma_i-{2+4k\over 3}\pi).
\eqno(7)
$$
Here $\rho$ and $\sigma$ are constant numbers; $k=1,\ldots,3$, the
addition in subscripts is understood modulo~3.

The method for calculation of vacuum curves is given in the work~[2].
As a result of direct calculations we get for the vacuum curve of
operator $L_i(\la)$ the equation
$$
1+\alpha_i(\la)u^3-\delta_i(\la)v^3-u^3v^3=0,
\eqno(8)
$$
where
$$
\alpha_i(\la)=-{a_i^3 \sin 3(\la+\rho_i)\over 4b_{21}^{(i)} b_{32}^{(i)}
b_{13}^{(i)}}; \quad
\delta_i(\la)=-{d_i^3 \sin 3(\la+\sigma_i)\over 4b_{21}^{(i)} b_{32}^{(i)}
b_{13}^{(i)}}.
\eqno(9)
$$

\oip{Lemma 2.} Two vacuum curves given by equations
$$
1+\alpha u^3-\delta v^3 -u^3 v^3=0
\eqno(10)
$$
and
$$
1+\alpha' u^3-\delta' v^3 -u^3 v^3=0
$$
commute if and only if $\alpha-\delta=\alpha'-\delta'$.

\medskip

{\it Proof } is obtained by straightforward calculation.

\oip{Theorem 3.} In order that two operator families
$L(\la)=L_1(\la)L_2(\la)$ and $M(\la)=L_2(\la)L_1(\la)$, where $L_1(\la)$
and $L_2(\la)$ are given by formulas (3--7), be equivalent it is necessary
and sufficient that the vacuum curves of operators $L_1(\la)$ and
$L_2(\la)$ commute for every~$\la$.

\section*{\S3. Proof of Theorem 3}

The necessity of the condition of Theorem~3 is obvious from Lemma~1.
The rest of this section is devoted to the proof of sufficiency.

We will need the following notations. We will always understand the operator
product of the type ${\cal L}=L_1\ldots L_k$ in the sense that those
operators have a common quantum space $S^Q$ and different auxiliary spaces
$S_1^A,\ldots,S_k^A$, so that $\cal L$ acts in $S^Q\otimes S_1^A\otimes
\cdots S_k^A$. Similarly to~(3), we will write $\cal L$ as ${\cal L}=
\pmatrix{{\cal A} & {\cal B} \cr {\cal C} & {\cal D}}$. Next, if
$L_i(\la)$ is an operator of the form (3--7) then we denote as
$L_i^{\dag}(\la)$ the operator of the same form given by the formula
$$
L_i^{\dag}(\la)=\pmatrix{0 & -1 \cr 1 & 0}\otimes \pmatrix{ && 1\cr
\bz & 1 & \cr 1 && \bz} \cdot L_i{\la}\cdot \pmatrix{0 & 1 \cr -1 & 0}
\otimes \pmatrix{ && 1\cr \bz & 1 & \cr 1 && \bz}.
$$
If $L_i(\la)$ and $L_i^{\dag}(\la)$ enter in the same product of operators
we will assume that they act in {\em different\/} copies of their
auxiliary space.

\oip{Lemma 3.} If the vacuum curve of operator $L_i(\la)$ has the equation
(8) then the vacuum curve of operator $L_i^{\dag}(\la)$ has the equation
$$
1-\delta_i(\la)u^3+\alpha_i(\la)v^3-u^3v^3=0.
$$
The composition of vacuum curves of operators $L_i(\la)$ and
$L_i^{\dag}(\la)$ has the equation
$$
(v^3-u^3)^3=0.
\eqno(11)
$$

\medskip

{\it Proof } is performed by direct calculations.

\medskip

Let us introduce the operation $\widehat{}$ of ``transposing in quantum
indices'': if $L=\pmatrix{A & B\cr C & D}$ then $\hat L=
\pmatrix{A & C\cr B & D}$.

\oip{Lemma 4.} If an operator $L_i(\la)$ is given by formulas (3--7)
then
\begin{eqnarray*}
&& \det L_i(\la)=[a_id_i\sin^2 \lambda -(b_{13}^{(i)})^2]^3,\\
&& \det \hat L_i(\la) = \det L_i(\la+{2\pi\over 3}).
\end{eqnarray*}

\medskip

{\it Proof } of the lemma is obtained by direct calculation.

\medskip

Thus, all zeros of functions $\det L_i(\la)$ and $\det \hat L_i(\la)$
are of order~3.

\oip{Theorem 4.} Let operators $L_1(\la)$ and $L_2(\la)$ be given by
formulas (3--7). Then for the operator
$$
{\cal L}(\la)=\pmatrix{{\cal A}(\la)&{\cal B}(\la)\cr
{\cal C}(\la)&{\cal D}(\la)}=L_1(\la)L_2(\la)L_2^{\dag}(\la)L_1^{\dag}(\la)
$$
there exists, in the general position in the parameters of operators
$L_1(\la)$ and $L_2(\la)$, the unique vector~$\Omega$ from the auxiliary
space such that ${\cal B}(\la)=0$ for all~$\la$. Vector~$\Omega$ is
an eigenvector for operators ${\cal A}(\la)$ and ${\cal D}(\la)$
with eigenvalues $a(\la)$ and $d(\la)$, and the function $d(\la)$
(resp.~$a(\la)$) has zeros of the first order in those and only those
points where $\det L_1 \det L_2 =0$ (resp.~$\det \hat L_1 \det \hat L_2 =0$).
The vacuum curve of operator ${\cal L}(\la)$ has the equation
$(v^3-u^3)^{27}=0$.

\medskip

{\it Proof. } According to Lemma~3, the vacuum curve of operator
$L_2(\la)L_2^{\dag}(\la)$ has the form~(11). It is easy to see that a curve
of the form~(11) commutes with any curve of the form~(8). Thus, the
vacuum curve of ${\cal L}(\la)$ coincides with the vacuum curve of
$L_1(\la)L_1^{\dag}(\la)L_2(\la)L_2^{\dag}(\la)$, i.e.\ with the composition
of two curves of the form~(11). This composition yields the curve
$(v^3-u^3)^{27}=0$.

It follows directly from the fact that the point $(u,v)=(0,0)$ belongs
to the vacuum curve of ${\cal L}(\la)$ for any $\la$ that, for any $\la$,
there exists such a vector $\Omega(\la)$ that ${\cal B}(\la)
\Omega(\la)=0$. It is known, however, that all operators ${\cal B}(\la)$
commute~[1]. So, the space $S^A$ where the operators ${\cal B}(\la)$ act
decomposes into a direct sum of subspaces $S_f^A$ corresponding to
different ``weights'' $f(\la)$ in the sense that the operators
${\cal B}(\la)-f(\la)\cdot{\bf 1}$ are nilpotent on~$S_f^A$. It is clear
 from the above that the function $f(\la)\equiv 0$ is necessarily present
among the ``weights''. In the corresponding subspace $S_0^A$ all the
${\cal B}(\la)$ are nilpotent and thus a vector $\Omega\in S_0^A$
exists such that ${\cal B}(\la)\Omega\equiv 0$.

Let us prove that the space of such vectors $\Omega$ is, in the general
position, not more than one-dimensional. If $b_{13}^{(1)}=b_{13}^{(2)}=0$
(formula~(6)) and the other parameters are in the general position then
necessarily
$$
\Omega={\rm const} \cdot \pmatrix{0\cr 0\cr 1} \otimes
\pmatrix{0\cr 0\cr 1} \otimes \pmatrix{0\cr 0\cr 1} \otimes
\pmatrix{0\cr 0\cr 1}.
$$
The space of vectors $\Omega$ cannot have a smaller dimension in a
particular case than in the general case.

The fact that $\Omega$ is an eigenvector for ${\cal A}(\la)$ and
${\cal D}(\la)$ follows now from the commutation relations of the
6-vertex model.

Clearly, the condition ${\cal D}(\la)\Omega=0$ together with
${\cal B}(\la)\Omega=0$
means the degeneracy of the operator ${\cal L}(\la)$; this can be achieved
only in the points where $\det L_1\cdot \det L_2=0$. Again from the case
$b_{13}^{(1)}=b_{13}^{(2)}=0$ we see that the function $d(\la)$ has
zeros of the first order in all such points. This statement is
extended to the
general case by continuity, taking into account that the whole number of
zeros of the function~$d(\la)$ cannot change~[4--7].

Similar arguments applied to the operator $\hat{\cal L}(\la)=
\hat L_1^{\dag}(\la) \hat L_2^{\dag}(\la) \hat L_2 (\la) \hat L_1 (\la)$
gives the theorem's statement as for function~$a(\la)$. The theorem
is proved.

\oip{Theorem 5.} Let, in addition to the conditions of Theorem~4,
the vacuum curves of operators $L_1(\la)$ and $L_2(\la)$ commute for
all~$\la$. Then the operator
$\tilde{\cal L}(\la)=\pmatrix{\tilde{\cal A}(\la) & \tilde{\cal B}(\la)\cr
\tilde{\cal C}(\la) & \tilde{\cal D}(\la)}=L_2(\la)L_1(\la)L_2^{\dag}(\la)
L_1^{\dag}(\la)$, too, possesses in the general position the unique vector
$\tilde\Omega$ such that $\tilde{\cal B}(\la)\tilde\Omega=0$,\quad
$\tilde{\cal A}(\la)\tilde\Omega=\tilde a(\la)\tilde\Omega$,\quad
$\tilde{\cal D}(\la)\tilde\Omega=\tilde d(\la)\tilde\Omega$,
and the zeros of $\tilde a(\la)$ and $\tilde d(\la)$ coincide, respectively,
with the zeros of $a(\la)$ and $d(\la)$. The vacuum curves of
$\tilde{\cal L}(\la)$ and ${\cal L}(\la)$ coincide.

\medskip

{\it Proof. } The vacuum curves of operators $L_1(\la)$ and $L_2(\la)$
commute, consequently the vacuum curve of $\tilde{\cal L}(\la)$ coincides
with the vacuum curve of ${\cal L}(\la)$. The further reasonings are
completely analogous to the proof of Theorem~4. The theorem is proved.

\medskip

It is clear from formulas (3--7) that $a(\la)$, $d(\la)$, $\tilde a(\la)$
and $\tilde d(\la)$ must be trigonometric polynomials. As the zeros of the
functions coincide as stated in Theorem~5, $\tilde a(\la)=a_0 a(\la)$ and
$\tilde d(\la)=d_0 d(\la)$, where $a_0$ and $d_0$ are constants. It follows
 from this and from the theory of commutation relations of the 6-vertex model
that the one-parametric family of operators $\tilde{\cal L}(\la)$ is
{\em equivalent\/} to the family
$$
{\cal M}(\la)=\pmatrix{a_0 {\cal A}(\la) & \root\of{a_0d_0}{\cal B}(\la)\cr
\root\of{a_0d_0}{\cal C}(\la) & d_0 {\cal D}(\la)}.
$$

At the same time, the vacuum curve equation for the operator
${\cal L}=\pmatrix{{\cal A} & {\cal B}\cr {\cal C} & {\cal D}}$ has
the form~[2]
$$
\det
\left[\pmatrix{1 & -v} \pmatrix{{\cal A} & {\cal B}\cr {\cal C} & {\cal D}}
\pmatrix{u\cr 1}\right]=0
\eqno(12)
$$
(here the operator in square brackets acts in the auxiliary space).
Thus, if the vacuum curve of operator ${\cal L}(\la)$ is given by the
equation ${\cal P}(u,v)=0$ then the vacuum curve of operator ${\cal M}(\la)$
must, as is easily seen, have the form
${\cal P}(\,\root\of{a_0\over d_0}u,\,\root\of{d_0\over a_0}v)=0$.
According to Theorems 4 and~5, both those curves have equations
$(v^3-u^3)^{27}=0$. Hence, $a_0=d_0$, and the operator family
$\tilde{\cal L}(\la)$ is equivalent to the family $a_0{\cal L}(\la)$.
We see from the case $L_2(\la)=L_1(\la)$ that $a_0\equiv 1$ (because
$a_0$ as a function of the parameters entering in families $L_1(\la)$
and $L_2(\la)$ cannot have jumps). Thus, we proved the following

\oip{Theorem 6.} If the vacuum curves of operators $L_1(\la)$ and
$L_2(\la)$ given by formulas (3--7) commute for all~$\la$ then the family
of operators $L_1(\la)L_2(\la)L_2^{\dag}(\la)L_1^{\dag}(\la)$ is
equivalent to the family $L_2(\la)L_1(\la)L_2^{\dag}(\la)L_1^{\dag}(\la)$.

\oip{Remark.} Similarly to Theorem~6, we can also show the family
$L_2^{\dag}(\la)L_1^{\dag}(\la)L_1(\la)L_2(\la)$ to be
equivalent to the family
$L_2^{\dag}(\la)L_1^{\dag}(\la)L_2(\la)L_1(\la)$.

\medskip

Consider now the following two operator families:
$$
{\cal L}(\la){\cal L}(\la)=L_1(\la)L_2(\la)L_2^{\dag}(\la)L_1^{\dag}(\la)
L_1(\la)L_2(\la)L_2^{\dag}(\la)L_1^{\dag}(\la),
\eqno(13)
$$
$$
{\cal L}(\la){\tilde {\cal L}}(\la)=
L_1(\la)L_2(\la)L_2^{\dag}(\la)L_1^{\dag}(\la)
L_2(\la)L_1(\la)L_2^{\dag}(\la)L_1^{\dag}(\la)
\eqno(14)
$$
(recall that different copies of operators act in different auxiliary
spaces). The families (13) and (14) have ``generating vectors''
$\Omega\otimes \Omega$ and $\Omega\otimes \tilde\Omega$ respectively,
so it follows from the theory of 6-vertex model's commutation relations that
if these two families are equivalent then the operator~$R$ performing this
equivalence is determined uniquely up to a numeric factor (in the general
position). At the same time, we get form Theorem~6 and the following Remark
{\em two\/} operators~$R$ acting in {\em different\/} auxiliary spaces
(and multiplied by $\bf 1$ in the rest, see Remark after Definition~2).
Thus, in reality, $R$ acts nontrivially only in the tensor product of
the auxiliary spaces of operators $L_1(\la)$ and $L_2(\la)$ that are
situated {\em on the right\/} in formulas (13) and (14); with this,
Theorem~3 is proved.

\section*{\S4. Discussion of results}

Formulas (8) and (9) show that the operator $R$ performing the equivalence
$$
RL_1(\la)L_2(\la)=L_2(\la)L_1(\la)R
\eqno(15)
$$
of operator families given by (3--7) exists if and only if the following
equality holds for all~$\la$:
$$
{a_1^3 \sin 3(\la+\rho_1)-d_1^3 \sin 3(\la+\sigma_1)\over b_{21}^{(1)}
b_{32}^{(1)} b_{13}^{(1)}} =
{a_2^3 \sin 3(\la+\rho_2)-d_2^3 \sin 3(\la+\sigma_2)\over b_{21}^{(2)}
b_{32}^{(2)} b_{13}^{(2)}}.
\eqno(15)
$$
At the same time, if the equivalence~(15) holds then vacuum curves of
the operator families $L_1(\mu+\la)$ and $L_2(\la)$,\quad $\mu={\rm const}$,
will {\em no longer\/} commute (except for some special cases).
Hence, due to Lemma~1, for such families the equivalence similar to~(15)
does not take place. This means that one cannot introduce
a parameter~$\mu$ similar to the parameter~$\la$ in $L_i(\la)$
in the set of all operators~$R$ performing equivalences of the type
considered in this paper. The author
hopes to investigate the issues of parameterization and the further
properties of operators~$R$ in the future. Here let us mention the two
following facts.

\oip{Theorem 7.} Let there be {\em three\/} operator families $L_i(\la)$,
$i=1,2,3$, given by formulas (3--7) and having mutually commuting vacuum
curves for all~$\la$. Let operators $R_{ij}$,\quad $i,j=1,2,3$,\quad
$i<j$, perform equivalences $R_{ij}L_i(\la)L_j(\la)=L_j(\la)L_i(\la)R_{ij}$.
Then the following equality (``triangle equation'') holds:
$$
R_{12}R_{13}R_{23}=R_{23}R_{13}R_{12}.
\eqno(17)
$$

\medskip

{\it Proof. } The left- and right-hand sides of equation~(17) both perform
the equivalence between operator families $L_1(\la)L_2(\la)L_3(\la)$ and
$L_3(\la)L_2(\la)L_1(\la)$. We can introduce three more operators and
auxiliary spaces and say that both sides of~(17) perform the equivalence
between families
$L_1^{\dag}(\la)L_2^{\dag}(\la)L_3^{\dag}(\la)L_1(\la)L_2(\la)L_3(\la)$ and
$L_1^{\dag}(\la)L_2^{\dag}(\la)L_3^{\dag}(\la)L_3(\la)L_2(\la)L_1(\la)$.
One can show in a way similar to the proofs of Theorems 4 and~5 that
the two latter families possess ``generating vectors'' and thus, according
to the theory of 6-vertex model's commutation relations, the operator
performing their equivalence is determined, in the general case, uniquely
(to within a constant factor). The theorem is proved.

\medskip

One more fact of interest is also that vacuum curves of the form~(10)
reveal obvious analogy to the vacuum curves of the well-known
Felderhof model studied for the first time in paper~[2].

\section*{References}

\frenchspacing

\begin{itemize}

\frenchspacing

\item[[1\hm]
Takhtajan L.A., Faddeev L.D. Quantum inverse problem method and the
$XYZ$ Heisenberg model. Uspekhi Mat. Nauk, V.~34, issue~5(209) (1979),
13--63 (in Russian).

\item[[2\hm]
Krichever I.M. Baxter equations and algebraic geometry.
Funkc. anal. i pril. V.~15, issue~2 (1981), 22-35 (in Russian).

\item[[3\hm]
Waerden B.L., van der. Algebra. Nauka Publishers, Moscow, 1979
(in Russian).

\item[[4\hm]
Korepin V.E. Analysis of the bilinear relation of the 6-vertex model.
Dokl. AN SSSR, V.~265, no.~6 (1982), 1361--1364
(in Russian).

\item[[5\hm]
Tarasov V.O. On the structure of quantum $L$-operators for the $R$-matrix
of the $XXZ$ model. Teor. Mat. Fiz., V.~61, no.~2 (1984), 163--173
(in Russian).

\item[[6\hm]
Tarasov V.O. Local Hamiltonians for integrable quantum models on a lattice.
II. Teor. Mat. Fiz., V.~61, no.~3 (1984), 387--392
(in Russian).

\item[[7\hm]
Izergin A.G., Korepin V.E. Lattice version of quantum field theory models
in two dimensions. Nucl. Phys., V.~B205[FS5], no.~3 (1982), 401--413.

\end{itemize}

\section*{Remarks of year 2000}

I hope that this paper is still of mathematical interest. On the historical
side, in order to prove who was really first in discovering some
facts related to the 6-vertex and chiral Potts models,
I would like to add here some more references, namely to papers
that I received back from journals in 1986--87. In particular, the same
as in this paper was already done not only for 3, but for any number of
``colors''. These are:

\begin{itemize}

\frenchspacing

\item
I.G. Korepanov, Vacuum curves of $\cal L$-operators associated with
the 6-vertex model, and construction of $R$-operators. Deposited at
VINITI (``All-Union Institute for Scientific and Technical Information'')
on April~2, 1986, Manuscript no.~2271-V86 (in Russian).

\item
I.G. Korepanov, Hidden symmetries of the 6-vertex model.
Deposited at VINITI on February~27, 1987, Manuscript no.~1472-V87
(in Russian).

\item
I.G. Korepanov, On the spectrum of the transfer matrix of 6-vertex model.
Deposited at VINITI on May~7, 1987, Manuscript no.~3268-V87
(in Russian).

\end{itemize}

Now this all (except some unimportant things) is published in the
following papers:

\begin{itemize}

\frenchspacing

\item
I.G. Korepanov, Vacuum curves of the $\cal L$-operators related to
the six-vertex model, St. Petersburg. Math. J., V.~6, no.~2 (1995),
349--364.

\item
I.G. Korepanov, Hidden symmetries of the 6-vertex model of
statystical physics, Zap. Nauch. Semin. POMI, V.~215 (1994),
163--177 (in Russian; English translation see in hep-th/9410066).

\end{itemize}

\end{document}